# Success in creative careers depends little on product quality

M.V. Simkin

Department of Electrical and Computer Engineering, University of California, Los Angeles, CA 90095-1594

## Abstract

In the recent article Janosov, Battiston, & Sinatra report that they separated the inputs of talent and luck in creative careers. They build on the previous work of Sinatra et al which introduced the Q-model. Under the model the popularity of different elements of culture is a product of two factors: a random factor and a Q-factor, or talent. The latter is fixed for an individual but randomly distributed among different people. This way they explain how some individuals can consistently produce high-impact work. They extract the Q-factors for different scientists, writers, and movie makers from statistical data on popularity of their work. However, in their article they reluctantly state that there is little correlation between popularity and quality ratings of of books and movies (correlation coefficients 0.022 and 0.15). I analyzed the data of the original Q-factor article and obtained a correlation between the citation-based Q-factor and Nobel Prize winning of merely 0.19. I also briefly review few other experiments that found a meager, sometimes even negative, correlation between popularity and quality of cultural products. I conclude that, if there is an ability associated with a high Q-factor it should be more of a marketing ability than an ability to produce a higher quality product.

Janosov, Battiston, & Sinatra [1] aimed to quantify the role of luck and individual ability in literary, music, and scientific careers. Their paper extends the work Sinatra et al [2] which was limited to scientific careers. A commonly accepted measure of success for scientists is the number of citations to their papers. This number varies wildly between different papers with most papers having just several citations and select few having thousands of those. Sinatra et al [2] observed that some scientists have a disproportional share of highly cited papers. They reasoned that this could not have happened if success in science was completely random. To explain this, they introduced what they called the Q-model. It states that the impact of the paper (they use the number of citations a paper gets during first 10 years since publication and denote it $c_{10}$ ) is $c_{10} = QP$. Here *P* is a lognormally distributed random variable and *Q* is fixed for each scientist but lognormally distributed among different scientists. The authors with high *Q* have many highly cited papers. The *P*-factor describes the role of randomness or luck and *Q*-factor describes talent or individual ability. Janosov et al [1] apply this Q-model to music, movies, and books. Instead of the number of citations they used the number of reviews, but mathematics is the same. The high-Q writers have a lot of popular books and so on.

There is, however, a problem. Yes, indeed, popular writers consistently produce popular books, but popular books do not get higher reader ratings than the unknown ones. In the Supplemental Information Section S1.3 [1] we learn that Spearman's rank correlation coefficient between the number of book ratings on Goodreads.com and the average rating is $r_s \approx 0.022$. Rank correlation coefficient between the number of movie ratings and the Metascore (average rating by movie critics) is higher, $r_s \approx 0.151$, but still low noting that $r^2 \approx 0.023$. They mention couple of other papers where "discrepancies between quality and popularity have been reported." However, there is another article where such discrepancy was indeed found though not clearly reported. It is [2], the original Q-factor article, the ideas of which [1] does develop.

## Citations and Nobel prize

Sinatra et al [2] computed the Q-factor for a large sample of scientists and ranked them accordingly. With the highest-Q scientist being number 1, second highest-Q - number two and so on. Afterward they studied how the Q-factor ranking relates to winning a Nobel Prize. They chose various rank thresholds and checked how many Nobel Prize winners and other scientists get in the sieve. They showed their results using a Receiver Operating Characteristic (ROC) in Figure 6(A) of [2]. This type of performance representation came from radar engineering. It is a two-dimensional plot of true positive (hit) versus false positive (false alarm) rates. In this case the true positive rate (TPR) is the ratio of the number of Nobel Prize winners in the sieve to the total number of them in the sample. False positive rate (FPR) is the ratio of other scientists in the sieve to the total number of them in the sample.

When you look at Figure 6(A) of [2] it looks like the classifier works well. We also see that the Q-factor is a better predictor of winning a Nobel Prize than the h-index or the total number of citations. This I will not contest. But let us look at the numbers. At certain rank threshold every Nobel Prize winner in the sample gets in the sieve. The TPR is 100%. The FPR at this rank threshold is merely 25%. But wait a second: Nobel winners are a tiny fraction of the scientists. So, there should be a lot more of them in the sieve even with the 25% FPR. The Q-factor is not that good a predictor after all.

To proceed we need to know $n_N$, the number of Nobel winners, and $n_O$, the number of other scientists in the sample. I could not find these numbers in Figure captions in [2]. I also could not get them from the authors. So, I had to extract the data from the plots using a plot digitizer. By matching the data from Fig. 6A and Fig. S45 of [2], I could compute the wanted numbers. My estimate is $n_N = 25$ and $n_O \approx 2890$. This means that at the rank threshold $R \approx 759$ when $TPR = 1$ and $FPR = 0.25$ we get 25 Nobel winners and 734 other scientists in the sieve. Note that if, for example, we had $n_O = 28$ and ROC plot was the same then at $TPR = 1$ and $FPR = 0.25$ we would get 25 Nobel winners and 7 other scientists in the sieve. This means that the ROC plot is not the best way to look at the data.

An alternative way to analyze the data is to compute a correlation coefficient. To each scientist we can attribute a Q-number which is 1 if they get in the sieve at given rank threshold and 0 otherwise. And an N-number which is 1 if the scientist got a Nobel Prize and 0 otherwise. We can interpret the N-number as a rating on a 2-point scale. Now we can compute a Pearson correlation coefficient between these vectors. A straightforward calculation gives:

$$r = \frac{(TPR-FPR)\sqrt{n_N n_O}}{\sqrt{(TPR \times n_N + FPR \times n_O)((1-TPR)n_N + (1-FPR)n_O)}} \qquad (1)$$

Note that Spearman correlation coefficient is equal to the Pearson one in this case under a mid-rank convention. At the rank threshold $R \approx 759$ when $TPR = 1$ and $FPR = 0.25$ Eq.(1) gives $r \approx 0.16$. This is in the same league with the low correlation coefficients in [1]. In the hypothetical case with $n_O = 28$ Eq.(1) would give $r \approx 0.77$. So, this method can distinguish between these two vastly different situations unlike the ROC plot.

Fig. S45 in [2] has the plot of the precision (the ratio of the number of Nobel winners in the sieve to the total number of scientists in the sieve) as a function of rank threshold. For $R = 1$ the precision is 0, meaning that the scientist with the highest Q-factor did not get a Nobel Prize. So, we have a negative correlation at this rank threshold. The next data point is at $R = 11$ where the precision is 0.27. This means that out of 11 highest Q-factor scientists 3 are Nobel winners. The correlation coefficient computed at this rank threshold using Eq.(1) is $r \approx 0.18$. It seems strange that the second data point is at the odd value of 11 rather than at a natural value of 10. My guess is that this is because precision reaches maximum at 11. This would imply that the second and third highest-Q scientists are not Nobel winners, for otherwise the maximum of precision would be at one of those thresholds. This would extend the range of negative correlation to two more rank thresholds: 2 and 3.

The maximum correlation ( $r \approx 0.27$ ) I found is for $R = 51$ when we get 10 Nobel winners and 41 other scientists in the sieve. This is better than what we got before but note that $r^2 \approx 0.07$ . We have different correlation coefficients for different rank thresholds. What is the right R to choose? I think it is $R = n_N = 25$. For than we get a perfect correlation ( $r = 1$ ) if the classifier works perfectly. In our case we get 5 Nobel winners and 20 other scientists in the sieve and $r \approx 0.19$.

I also estimated the Q-ranks of all 25 Nobel winners in the dataset of [2] and computed Spearman rank correlation coefficient with Nobel-rank under a mid-rank convention. Under this convention each of 25 Nobel winners got the rank 12.5 and each 2890 non-winner the rank $12.5 + 2890/2 = 1457.5$. I got $r_s \approx 0.12$ . However, this is not the right way to analyze the data since even if all 25 Nobel winners got the highest Q-rank we would still get $r_s \approx 0.21$ .

I conclude that the ROC plot in Figure 6(A) of [2] overrepresents the accuracy of the classifier. It is not the right way to look at the data when the ratio of positives and negatives in the sample is so small. The correlation between the Q-factor and Nobel prize winning is on par with the low correlations between fame and quality ratings for books and movies reported in [1].

## More data on quality versus success

First, I would like to analyze the classic study [3] done in 1920s by I.A. Richards, a lecturer in English at Cambridge. He gave his students, most of whom were English majors with a view to an Honours Degree, a set of 13 poems by 13 different poets and asked to judge their worth. The trick was that he withdrew the names of the authors. Famous poets did worse than the unknown ones when their names were detached from their poetry. For example, H. W. Longfellow got only 5% of positive ratings.

To study quantitatively the relation between quality rating and fame I measured the number of webpages mentioning each poet using Google. I used full poet's name as an exact phrase plus a separate word "poet" in Google searches. In Figure 1(a) you can see the scatter plot of the fraction of positive ratings versus the number of Google hits. The fame decreases as the rating increases. The Spearman rank correlation coefficient between Google hits and rating is $r_s \approx -0.71$.

I also looked up the numbers of different books by the poets in question using WorldCat [4] in libraries worldwide. A plot of the ratings versus the number of WorldCat hits, looks so much like Fig. 2(a) that there is no need to add it. The Spearman rank correlation coefficient between WorldCat hits and rating is $r_s \approx -0.76$. In contrast the correlation between the two different measures of fame (Fig. 2(b)) is highly positive. The Spearman rank correlation coefficient between WorldCat hits and Google hits is $r_s \approx 0.90$. High correlation between different measures of fame is a common thing: see Figure S2 of [1] and Fig. 2 of [5].

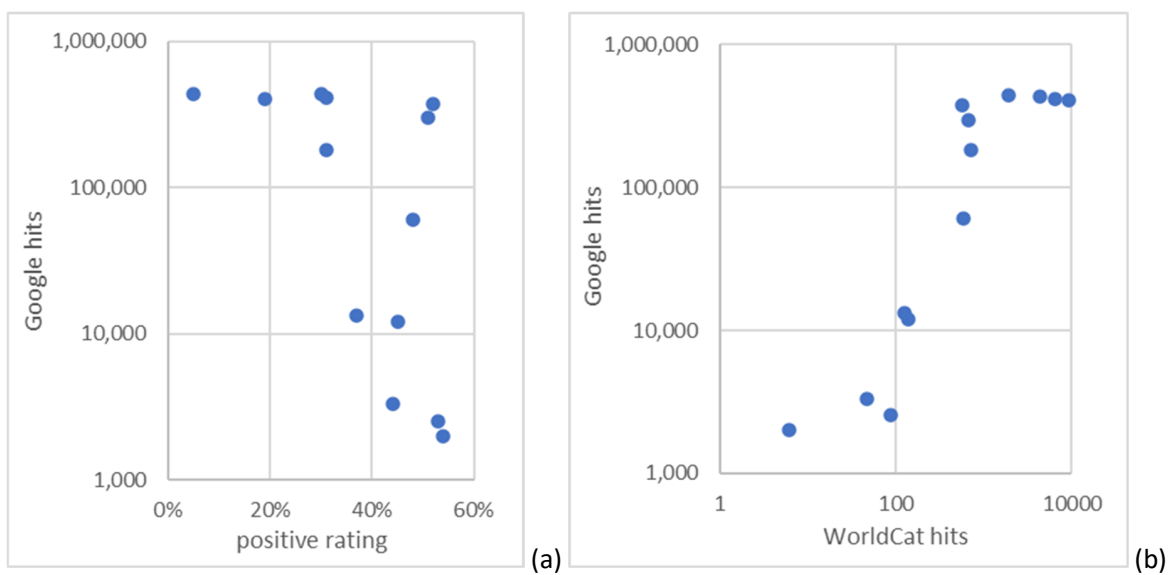

*Figure 1.* **(a)** The fraction of positive ratings in Richards' experiment versus the number of webpages mentioning the poet in question that I found using Google. **(b)** Number of books by poets in question in libraries worldwide I found using WorldCat versus the number of Google hits. Note that I produced this figures in April 2017. Today's numbers of Google and WorldCat hits are different.

The first possible explanation for this strongly negative correlation between quality and fame is that the sample is small. Using Fisher's r-to-z transform [6] we can estimate the confidence interval for $r_s$. For the 95% confidence interval the upper limit is $r_s^+ \approx -0.27$. For the 99% we get $r_s^+ \approx -0.080$. So, the correlation is clearly negative though it could be a lot smaller in absolute value.

This is at odds with the small but positive correlation $r_s \approx 0.022$ [1] between the average Godreads book rating and the number of those ratings. Because of millions of data points in the set we can only be uncertain of the second significant digit. A possible explanation of this disparity is that Richards selected the best poems of the unknown poets and the worst poems of the famous poets. Another possible explanation is that the participants of Richards' experiment did not know the names of the poets while Goodreads raters knew the names of the book authors.

In one blind wine testing experiment [7] which included over 6,000 trials the correlation between wine price and its average taster's rating was also negative. I am not sure what the correlation coefficient exactly is since Goldstein et al [7] report their results in the form of linear regression coefficient between average rating and the natural logarithm of wine price. This coefficient is $b_1 \approx -0.04$. One can express Pearson correlation coefficient through the regression coefficient, but one needs to know sample variances to do this. And they are not reported. Fortunately, we know the ranges. Taster's rating is on 4-point scale, so the difference between the extremes is 3. The price of wine ranged between \$1.65 and \$150. So, the difference between the extremes is $\ln \frac{150}{1.65} \approx 4.5$. If sample variances scale roughly as the squares of the ranges, we get for the Pearson correlation coefficient between rating and logarithm of price is $r \approx b_1 \frac{3}{4.5} \approx -0.027$. This assumption may be too strong, but the number should be correct at least by the order of magnitude.

There also have been two earlier studies [8], [9] which just like [1] had compared the number of Goodread ratings with the average rating. Simkin [8] compared ten most rated books of Charles Dickens and Edward Bulwer-Lytton. Dickens had thousand times more ratings but only 6% higher average rating. Fisher [9] did a similar comparison of Virginia Woolf and Arnold Bennett. Woolf had tens of times more ratings but only 5% higher average rating than Bennet. However, both studies found a larger difference between the authors when one compared not the average ratings, but the percentages of top ratings. So, I think that if one would redo the analysis of [1] using percentages of top ratings instead of average ratings one woul get a higher correlation coefficient. Albeit being higher it still is not going to be high.

In one internet experiment [10] over 9,000 people tried to tell the prose of Charles Dickens from that of Edward Bulwer-Lytton. The result was on the level of random guessing. The subset of 76 participants from Ivy League and Oxbridge did on the same level as the crowd.

Ross [11] retyped a novel by Jerzy Kosinski that won a National Book Award. Submitted it under a fictitious name to 14 publishers and 13 agents. No one recognized the work, and nobody thought it deserved publishing. Including the original publisher. Three similar experiments [12], [13], [14] with fiction publishers yielded similar results.

I could go on with many more relevant experiments in writing alone. Set aside visual arts and music. But this would be too long for a commentary.

## Discussion

The presented above facts should leave little doubts that the small correlation coefficients between fame and product quality mean anything else than the actual lack of such correlation. The argument [1] that correlation coefficients could be low because performance and popularity follow different probability density distributions is not appealing. If you use a Pearson correlation coefficient you can take a logarithm of popularity to adopt to its heavy-tailed nature and increase the correlation coefficient as was done in [4]. For Spearman rank-correlation coefficient you do not even need to do such transformation since it will not change the result. If there is a correlation – Spearman coefficient will show it.

Janosov, Battiston, & Sinatra [1] claim that they separated luck and individual ability in creative careers. This for the moment I shall not contest (I will do this in a separate article). Indeed, some people have a high Q-factor and consistently produce more popular works. But this more popular work is not of a significantly higher quality than the work of the low-Q individuals as the low correlation coefficients between fame and quality have demonstrated. So, if there is an ability linked with a high Q-factor it should be more of a marketing ability than an ability to produce a product of higher quality.